\documentclass[12pt,preprint]{aastex}
\usepackage{graphics,lscape,ccaption}
\usepackage{amsmath,bm}
\usepackage{times}
\usepackage{epstopdf}
\begin{document}
\slugcomment{Submitted to ApJ}

\title{White-Light Flares on Close Binaries Observed with {\it Kepler}}
\author{Qing Gao\altaffilmark{1,2}, Yu Xin\altaffilmark{1}, Ji-Feng Liu\altaffilmark{1}, 
Xiao-Bin Zhang\altaffilmark{1}, Shuang Gao\altaffilmark{3}}

\altaffiltext{1}{Key Laboratory of Optical Astronomy, National Astronomical Observatories,
Chinese Academy of Sciences, Beijing 100012, China;~~\it{gaoqing10@mails.gucas.ac.cn}}
\altaffiltext{2}{University of Chinese Academy of Sciences, Beijing 100049, China}
\altaffiltext{3}{Department of Astronomy, Beijing Normal University, Beijing 100875, China}

\begin{abstract}
Based on $\it{Kepler}$ data, we present the results of a search for white-light flares 
on \textbf{1049} close binaries. We identify \textbf{234} flare binaries, 
on which \textbf{6818} flares are detected.
We compare the flare-binary fraction in different binary morphologies (``detachedness''). 
The result shows that the fractions in over-contact and ellipsoidal binaries are approximately 
10-20\% lower than those in detached and semi-detached systems.
We calculate the binary flares activity level ($AL$) of all the flare binaries, and discuss its 
variations along the orbital period ($P_{\rm{orb}}$) 
and rotation period ($P_{\rm{rot}}$, calculated for only detached binaries).
We find that $AL$ increases with decreasing $P_{\rm{orb}}$ or $P_{\rm{rot}}$ up to the critical values 
at $P_{\rm{orb}}\sim$3~days or $P_{\rm{rot}}\sim$1.5~days, thereafter, the $AL$ starts decreasing 
no matter how fast the stars rotate.
We examine the flaring rate as a function of orbital phase in 2 eclipsing binaries 
on which a large number of flares are detected.
It appears that there is no correlation between flaring rate and orbital phase in these 2 binaries. 
In contrast, when we examine the function with \textbf{203} flares on \textbf{20} non-eclipse ellipsoidal 
binaries, bimodal distribution of amplitude weighted flare numbers shows up at orbital phase 
0.25 and 0.75.
\textbf{Such variation could be larger than what is expected from the cross-section modification.}

\end{abstract}

\section{INTRODUCTION}
Flares are short-lived and intense release of magnetic energy caused by the 
reconnection of magnetic field loops in the outer atmosphere of stars 
(e.g., \citet{Benz2010,Shibayama2013}).
The typical energy release in a solar flare ranges from $10^{27}$ to $10^{32}$
erg, and the duration is several minutes to several hours. 
Ground-based observations show that stellar flares occur most frequently in M dwarfs and 
are typically 10-1000 times more energetic than solar flares (e.g., \citet{Gudel2009}). 
Nine superflare candidates on G-type dwarfs with a typical energy of 
$10^{33}-10^{38}$ erg have been reported by \citet{Schaefer2000}. 
Such search was extremely extended by \citet{Shibayama2013}, in which 1547 superflares 
are found on 279 G-type dwarfs from {\it Kepler}.

The {\it Kepler} spacecraft is quite powerful in detecting stellar flares due to its 
high-precision ($\sim$~0.1~mmag) and long-period time-series observations \citep{Koch2010}. 
Since its launch, the number of flare stars in variety of stellar types has been 
dramatically increased, for instance, 373 flaring stars among $\sim$ 23,000 cool dwarfs 
are discovered in the {\it Kepler} Quarter 1 long cadence data \citep{Walkowicz2011}, 
flare A-type stars have been detected in the {\it Kepler} field \citep{Balona2013}, 
and 148 G-type superflare stars are found from $\sim$ 83,000 stars observed over 120~days 
\citep{Maehara2012}. 
Meanwhile, our knowledge of the typical energies of stellar flares has been changed.
Most of the flares caught with {\it Kepler} are $10^2-10^6$ more intense than the 
large solar flares, which should be identified as `superflares' if from 
the ground-based observations. 

Flares were also discovered on binaries, e.g., 
RS~CVn variables \citep{Pandey2012} and W~UMa star (CSTAR~038663; \citet{Qian2014}). 
RS~CVn is a class of detached binaries typically composed of a chromospherically 
active G or K star (e.g., \citet{Berdyugina2005}). 
The underlying mechanisms of triggering flares on RS~CVn are suggested to remain 
the same as in the single stars, the magnetic reconnection in the field connecting the 
two stars may cause the flare events more active (e.g., \citet{Gunn1997}), 
\textbf{ and tidal effect plays important role for enhancing magnetic activity of
RS~CVn stars \citep{Cuntz2000}.}
Flares on CSTAR~038663 are suggested to be originated on the surface of the contact binaries 
based on the observational facts that $(i)$~the flare amplitude is increasing along the 
duration, and $(ii)$~the photometric solution reveals a long-lived dark spot on 
the contact binary, which could be associated with the flares \citep{Qian2014}.
Using {\it Kepler} data, \citet{Balona2015} examined the flare frequency as a function of 
orbital phase in three eclipsing binaries. Unfortunately, no correlation of flaring with orbital 
phase was established from the work, and the result weakens the hypothesis that flares in 
close binaries could be a result of reconnection of field lines connecting the two stars 
\citep{Balona2015}. 

The main goal of this work is to identify flare binaries from {\it Kepler} data, and to investigate 
the correlations between flare properties and the physical parameters of binary systems. 
In order to achieve the purpose, we analyzed the light curves of \textbf{1049} close binaries, identified 
\textbf{234} flare binaries, and discovered \textbf{6818} flares on them. 
In Section~2, we describe the criteria of working-sample selection and the method 
of flare identification. In Section~3, we discuss the flaring properties of two `peculiar' 
systems in Section~3.1, i.e., Kepler-Input-Catalog~(KIC)~5952403 (a triple star system) and 
KIC~11347875 (a binary with asymmetric light curve), 
we detect flaring rate along the flare intensity and binary morphology in Section~3.2, 
along the orbital period and rotation in Section~3.3, and along the orbital phase in Section~3.4. 
The conclusion and discussion are given in Section~4. 

\section{Observation and Data Analysis}
\subsection{{\it Kepler} Data}
The {\it Kepler} spacecraft was launched by NASA in March 2009 with the primary objective of 
searching for exoplanets by finding planetary transit events from the recorded light curves of 
target stars in the Cygnus, Lyra, and Draco region \citep{Borucki2011b}. As the luminosity decrease 
from planetary transit is usually less than one-hundredth of the total brightness of the 
star and the planet can be observed in front of the star only when the orbit is nearly 
parallel with the line of sight, {\it Kepler} is thus designed to obtain high-precision and 
long-term light curves of over 190,000 stars.
The typical photometry precision is 0.1~mmag for a star of 12~magnitude, 
and the photometry was continuously processed and recorded for a period of four years
\citep{Koch2010,Gilliland2010a}. 
Therefore, the {\it Kepler} light curves are useful in detecting not only planetary transits but also 
other small brightness variations such as stellar flares. 

{\it Kepler} has two kinds of data records which are specified with their time resolutions, i.e., 
long cadence (LC) of 29.4 minutes exposure and short cadence (SC) of 1 minute exposure.
Flare detection from {\it Kepler} is confined to the LC exposure time. 
The 30-min exposure prohibits the detection of short-lived flares and greatly 
lowers the visibility of flares lasting less than one or two hours. 
SC exposure is more suitable for the study of stellar flares \textbf{\citep{Maehara2015}}. 
Unfortunately, SC data are available for only 4828 stars and the time coverage is usually 
less than two months \citep{Gilliland2010b}.

\subsection{Sample Selection}
{\it Kepler} light curves are released in two types, i.e., the original simple aperture 
photometry, and with the pre-search data conditioning (PDC). 
The instrumental effects and some other errors such as flux discontinuities are removed 
from the PDC data \citep{Smith2012}.
As the first step of the work is to identify flare binaries as many as possible, 
a sufficient working sample in accurate photometry can certainly help making 
the statistical results more reliable, therefore, we choose the {\it Kepler} 
LC-PDC data in the work {\footnote{\textbf{There are different release version of the 
{\it Kepler} LC-PDC data (http://archive.stsci.edu/kepler/data\_release.html). 
The version we adopted for each flare binary is listed in column~11 in Table~1.}}}.

The {\it Kepler} eclipsing binary (KEB) 
catalog\footnote{\textbf{$\bf{http://keplerebs.villanova.edu/}$}} 
\textbf {(released on June 6, 2015 \citep{Kirk2015})}
includes more than 2800 binary stars identified from {\it Kepler} data. 
KEB provides the binary information of its KIC identifier, 
morphology type, binary period, coordinates, {\it Kepler} magnitude, and etc, among which, 
the morphology is valued from 0 to 1 matching with the binary type from detached (0$\sim$0.5), 
semi-detached (0.5$\sim$0.7), over-contact (0.7$\sim$0.8), to the ellipsoidal (0.8$\sim$1.0) 
binaries, using the method of locally linear embedding \citep{Matijevic2012}.

We select \textbf{1049} out of the over 2800 binaries in KEB as our working sample,
based on mainly three criteria: \textbf{(1) binaries with neighboring stars within 12 arcsec 
are excluded to avoid false events from the flux of the neighboring stars (\citet{Shibayama2013})};
(2) the orbital period is longer than 0.4 days, since the light 
curves with period less than 0.4 days are easy to be entangled with the flare profiles; 
and (3) as we use the 30-min exposure LC data, given the period of $\geq$~0.4~days, 
there has to be more than 20 data records within one orbital period, 
otherwise, the decomposition of flares from the binary light variation 
may become impossible. \textbf {Meanwhile, some `peculiar' targets are also 
excluded from the working sample, which are mainly: (1) 42 heartbeat binaries} 
with highly eccentric orbit and tidally induced pulsations, as their light curves are unable to be 
understood with 4-terms Fourier function which we adopt in the work for flare 
detection (detailed description of the method is given in Section 2.3); 
\textbf {(2) 88 targets being not binaries, i.e., single stars with stable dark spots 
\footnote{\textbf{The stable dark spots rotate together with the single star, which makes the 
light curve of the single star very easily miss-classified as that of the ellipsoidal binary. 
It can be distinguished from such as the flip-flop phenomenon (Jetsu et al. 1994; Jetsu 1996) 
and the amplitude diminishing on the light curve 
due to the weakening or vanish of the dark spots.}} 
and multiple-star systems; 
and (3) 17 KIC targets without valid morphology values.}

\subsection{Flare Measurements}

The flare detection efficiency is limited by the data mode and the photometric uncertainty.
Besides the disadvantage of using the long-time-exposure LC data, 
there is another difficulty that we need to confront at first, i.e., as the flare detection 
is based on the measurement of the light curve variations, 
a detrend process is definitely required to eliminate 
the baseline flux off the light curve.  

Given that the ellipsoidal binaries have the most smooth binary light curves that present the 
quasi-sinusoidal variations, and the variation can be well understood with 
no more than 3 terms Fourier series \citep{Morris1985}, we decide to use 4 terms Fourier 
series to trace all the binary light curves without any smoothness 
except for the detached systems, as we are working on a large binary sample 
including different morphology types.

As for the detached binaries, 
the two components have no interaction or mass exchange between each other, 
and thus they can be considered as two `isolated' single stars;
therefore, the baseline fitting of detached binaries follows the method 
adopted in the work of flare detection on cool single stars by \citet{Walkowicz2011}, 
i.e., median filter the observed light curves on which 
the flares have been tagged out and then create the baseline curve with quiescent 
variability but no flares. The data points during eclipse are excluded.

The main procedure of baseline fitting for the other three morphology types 
follows the steps as:

\begin{itemize}

\item[$(i)$] The observed light curve is fitted with 4 terms Fourier series in each orbital period. 
The data that are $3\sigma$ away from the Fourier fitting curve will be eliminated. 
The process iterates until all the data left are within $3\sigma$.

\item[$(ii)$] The remaining data after step~$(i)$ are fitted again with 4 terms Fourier series. 
This time, the fitting is repeated 10 times with 0.1 orbital period backwards each time, 
which means each observing data point refers to 10 fitting points.

\item[$(iii)$] The average and standard deviation of the 10 fitting points are calculated
for each data point. Any fitting points $1\sigma$ away from the average are discarded.

\item[$(iv)$] Finally, the baseline curve is established by taking the $\sigma$-clipped 
robust mean value of the fitting points left after step~$(iii)$.

\item[$(v)$] When the method is applied to the semi-detached and over-contact binaries, 
the data points during eclipsing are simply excluded.

\end{itemize}

\begin{figure}[t]
\includegraphics{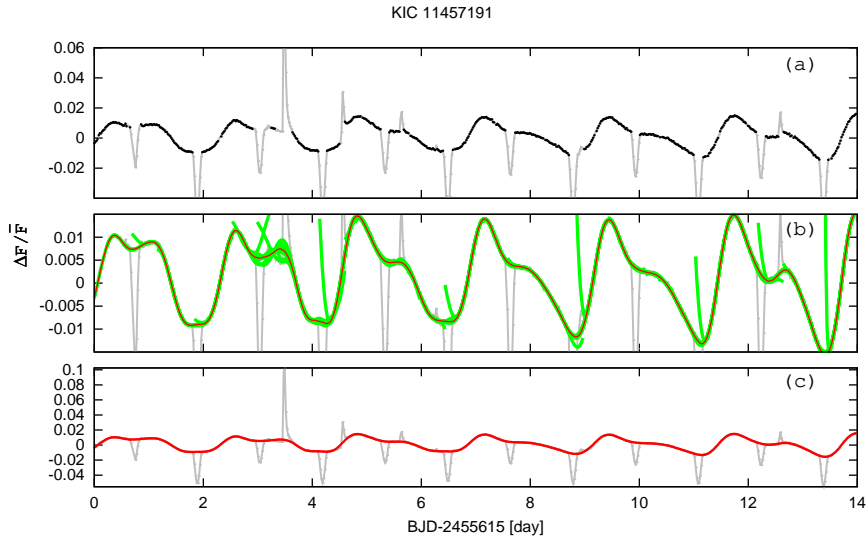}
\caption{\label{fig1} \textbf{Procedure of the baseline fitting. 
The light curve of KIC~11457191 is used as an example. 
Panel~(a): The original light curve is plotted in grey line. The black dots 
are the remaining data points after the $3\sigma$ exclusion in step $(i)$ (Section~2.3).
As KIC~11457191 is a semi-detached binary, data points during eclipsing are also 
excluded. Panel~(b): 10 times fitting points in step $(ii)$ are given in green lines. The 
outliers clearly show up around the green-line discontinuities.
Panel~(c): The baseline curve (red line) is established by connecting the 
$\sigma$-clipped robust mean value of the fitting points (the green line in panel~(b)). }}
\end{figure}

\textbf{Figure.~\ref{fig1} demonstrates the procedure of baseline fitting. The light curve 
of KIC~11457191, a semi-detached binary, is used as an example. 
The original light curve is plotted in grey lines in the figure. 
The remaining data for baseline fitting after $3\sigma$ exclusion 
in step $(i)$ are plotted as black points in panel~(a). 10 times fitting points in step $(ii)$ are 
given in green lines in panel~(b), where the outliers clearly show up around the green-line 
discontinuities. The baseline curve (red line in panel~(c)) is established by connecting the 
$\sigma$-clipped robust mean value of the fitting points (the green line). 
Panel~(c) presents the original light curve (grey line) and the final baseline 
curve (red line) of the target.}

There are various flare detection algorithms (e.g., \citet{Walkowicz2011,Osten2012}).
As a general description, light curves are analyzed after they have detrended, i.e., 
the baseline curve has been subtracted from the observed light curve,
and are searched for brightness enhancement where the relative flux becomes 
statistically larger than a certain threshold for two or more times consecutively. 
In this work, we combine the algorithms of \citet{Walkowicz2011} and \citet{Shibayama2013} 
to identify flare binaries:

\begin{itemize}

\item[$(i)$] Events are flagged as flare candidates when a minimum of 3 contiguous points 
are found above a threshold of 3 times the standard deviation of the local quiet flux.

\item[$(ii)$] We calculate the number distribution of brightness variation between 
all pairs of two consecutive data points for all the binary stars.  

\item[$(iii)$] Brightness variation from step $(ii)$ is identified as flare when 
the event is marked as flare candidate from step $(i)$ and  
it matches the threshold of the `1\%' number distribution from step~$(ii)$ 
(as presented in Figure.~1(b) in \citet{Shibayama2013}). 

\item[$(iv)$] \textbf{Pairs of flares that occurred at the same time and whose 
spatial distance is less than 24 arcsec are excluded to avoid mis-detection 
due to the contaminations of other sources or the instrumental uncertainties 
such as the variation of pointing accuracy and cosmic rays (\citet{Shibayama2013}).}

\item[$(v)$] \textbf{Light curves after step~$(iii)$ and step~$(iv)$ are confirmed by eye. 
The step is necessary to mainly reduce the false flares due to cosmic-rays and 
{\it Kepler} artifacts such as focus changes and loss of fine point
\footnote{\textbf{http://archive.stsci.edu/kepler/manuals/KSCI-19033-001.pdf}}, 
which may cause discontinuities in the light curves. 
The features might be captured as flares by computer, but easily distinguished by eye. 
There are approximately 9\% false flares rejected in this process.}

\item[$(vi)$] \textbf{As a final step, the pixel level data of each remaining flare from step~$(v)$
are examined by eye. If the spatial distribution of a flare on the CCDs is different from the 
distribution of the most energetic quiescence, the flare is considered as a brightness variation of 
another source (\citet{Shibayama2013,Kitze2014}). 
About 12\% of flare candidates are removed in this step.}

\end{itemize}

\section{Data Results and Analysis}

Following the method described in previous section, we identify \textbf{234} flare binaries from the
\textbf{1049} close binary samples.  \textbf{6818} flares are detected on flare binaries. 
Fundamental parameters of the \textbf{234} flare targets and the \textbf{6818} flares 
(together with light curves)  are available electronically. 
Table~1 and Table~2 give the excerpt from the overall tables.

\begin{figure}[t]
\includegraphics{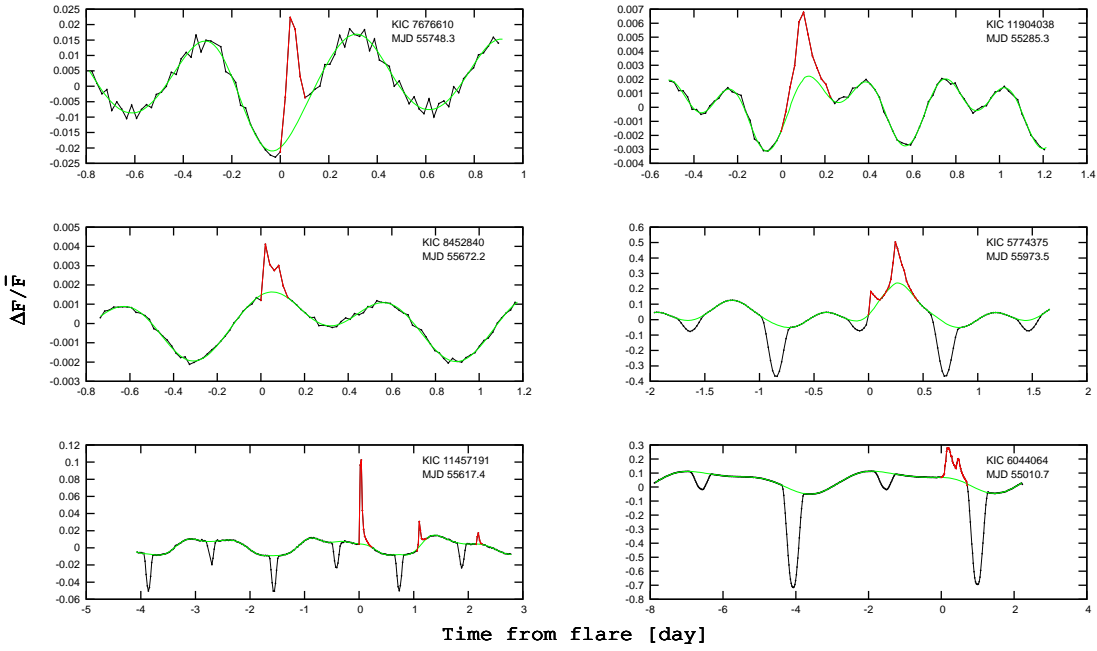}
\caption{\label{fig2} Light curves of flares on 6 close binaries. Horizontal and vertical 
axes correspond to the time (day) from the flare, and the relative flux ($\Delta F/\bar F$), respectively. 
In each panel, the black, green and red lines present the light curves of the observed data,
the baseline fitting flux, and the flares, respectively. 
The KIC number and the modified julian date (MJD) of flare peak time of each binary are 
given in the upper right corner in each panel.}
\end{figure}

Figure.~\ref{fig2} shows the examples of the flare light curves on 6 close binaries.
Horizontal and vertical axes correspond to the time (day) from the flare, and the relative flux 
($\Delta F/\bar F$) from eq.~(1), respectively. 
In each panel in Figure.~\ref{fig2}, the black, green and red lines represent the light curves of 
the observed data, the baseline fitting flux, and the flares, respectively. 
The KIC number of each binary is given in the upper right corner in each panel.
\begin{equation}
\Delta F/ \bar F = (F - \bar F) / \bar F, 
\end{equation}
where $F$ is the observed flux from LC-PDC data. 
$\bar F$ is the average flux during the current observation quarter. 

Based on the light curves, there are basically two detectable parameters 
regarding flare properties.
One is the total number of flares ($N_{\rm f}$) during the entire observing period, the other 
is the corresponding flare intensity. $N_{\rm f}$ is directly obtained from flare detection. 
The measurement of flare intensity is processed with equations (2) and (3).

For each flare, we integrate the points that are tagged as part of the flare, 
which is essentially a photometric equivalent width ($EW$) of the flare \textbf{\citep{Walkowicz2011}}, i.e.,

\begin{equation}
EW=\int \frac{F_{\rm f}-F_{\rm q}}{F_{\rm q}} dt,
\end{equation}
where $F_{\rm f}$ and $F_{\rm q}$ are the flaring and the quiescent flux, respectively. 
\textbf{The observed flare energy is interpolated with the cubic spline function for $EW$ integral.
The integral starts and ends at the baseline curve, and the interpolation usually includes 1000 grids.
We compare the $EW$ results when performing the integral with or without spline interpolation. 
Two flares, one with 1.5-hour duration and 6 data points, the other with 13-hour duration and 
29 data points, are used as the sample. The results show that: 
(1) for the 6-point flare, the $EW$ difference is 9.6\%; 
and (2) for the 29-point flare, the difference is 1.1\%. 
As we mentioned in previous content, the SC data are actually more suitable for the studies of stellar 
flares compared to the LC data \citep{Maehara2015}. \citet{Shibayama2013} showed two 
examples of the flare energy difference from using SC and LC data, there is no energy difference 
for the flare in Figure.~11, and the SC result is 20\% larger than the LC result in Figure.~12 
(See appendix A in their paper). 
Considering that flares in SC data are sampled more sufficiently, and thus should present higher 
(at least equal) flare energy compared to those in LC data, 
the cubic spline interpolation we adopt in the 
$EW$ integral will not artificially expand the flare intensity and will decrease the possibility of 
underestimating the $EW$.}

Such an $EW$ quantity has actually the units of time. It can be intuitively thought of as the time 
interval over which the quiescent star emits as much energy as was released during of 
the flare. Meanwhile, $EW$ is a differential quantity, independent of distance and measured 
relative to the quiescent star. Therefore, for one binary system, the summation of $EWs$ 
of all the flares divided by the total observation time can provide a measurement of the 
flaring intensity, which is \textbf{defined} as the flare `activity level
\footnote{\textbf{To our knowledge, the concept of `activity level' ($AL$) is used for the first time 
to estimate the ratio of flare energy over quiescent radiation energy.}}' ($AL$) in the work:

\begin{equation}
AL=\frac{\sum\limits_{i=1}^{N_{\rm f}} EW_{\rm i}}{T_{\rm{obs}}} \approx \frac{E_{\rm f}}{E_{\rm q}},
\end{equation}
where $EW_{\rm i}$ is the $EW$ of the $ith$ observed flare on a binary. 
$T_{\rm{obs}}$ is the entire observation time of the binary. 
Assuming that the binary quiescent radiation is relatively a stable value, 
i.e., $F_{\rm q}$ in eq.~(2) is approximately a constant, the description of ($F_{\rm q}\times T_{\rm{obs}}$) 
represents thus the total energy of the quiescent radiation ($E_q$), the value of 
($F_{\rm q}\times\sum\limits_{i=1}^{N_{\rm f}}EW_{\rm i}$) is the total energy of all the flares ($E_{\rm f}$), 
and therefore the $AL$ value actually estimates the ratio of $E_{\rm f}/E_{\rm q}$.

\begin{figure}[t]
\includegraphics{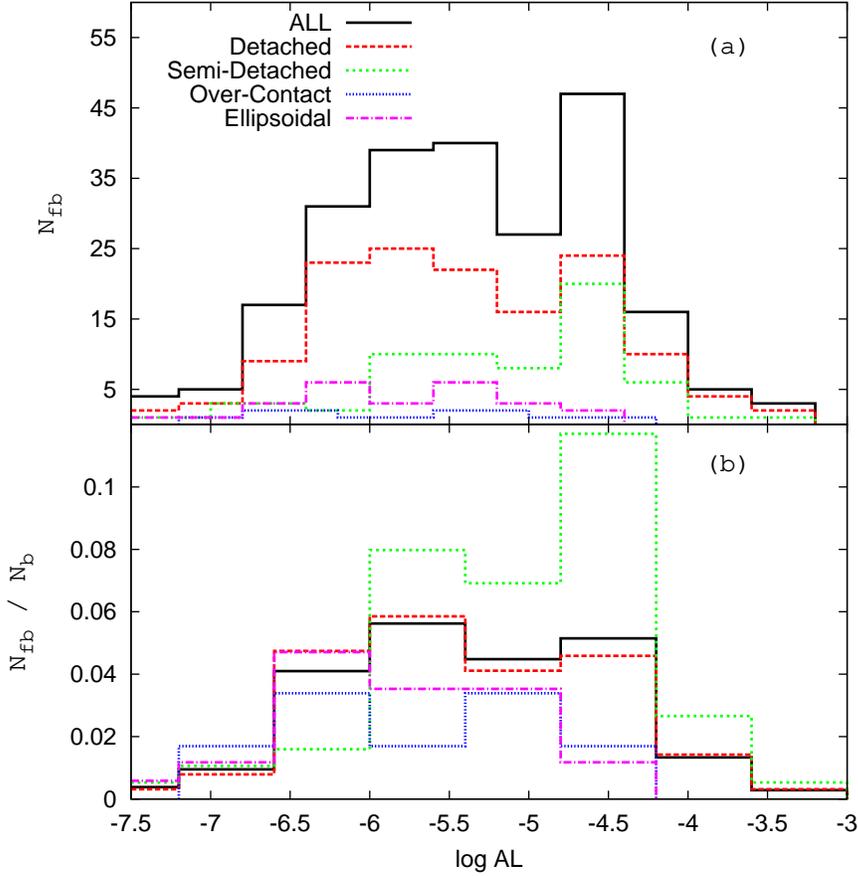}
\caption{\label{fig3} Panel~(a): Number distribution of flare binaries ($N_{\rm{fb}}$) along the 
$log~AL$ values. The black, red, green, blue and magenta lines show the distribution of the total 
234, the detached, semi-detached, over-contact and ellipsoidal flare binaries, respectively. 
The two peculiar binaries discussed in Section 3.1 are excluded. 
\textbf{Panel~(b): Distribution of $N_{\rm{fb}}/N_{\rm{b}}$ ratio along $log~AL$. 
As $AL$ can only be calculated for flare binaries, $N_{\rm b}$ is distinguished only 
for binary types when calculating $N_{\rm{fb}}/N_{\rm{b}}$ in different $log~AL$ bin.}}
\end{figure}

Figure.~\ref{fig3}(a) presents the number distribution of flare binaries ($N_{\rm{fb}}$) along the 
$log~AL$ values.
The black line in the figure represents the $N_{\rm{fb}}$ distribution of the total \textbf{234} flare binaries. 
The red, green, blue and magenta lines are the $N_{\rm{fb}}$ distributions of the 
detached, semi-detached, over-contact and ellipsoidal flare binaries, respectively.
The black line shows two peaks at $log~AL$ around -4.5 and -5.5.  
The `-4.5' peak is mainly contributed by the detached (red line) and semi-detached (green line) 
systems, \textbf{which indicates more energetic flares occur on these two types compared to
the over-contact and ellipsoidal systems.}
\textbf{Figure.~\ref{fig3}(b) presents the distribution of $N_{\rm{fb}}/N_{\rm{b}}$ along $log~AL$. 
As $AL$ can only be calculated for flare binaries, $N_{\rm b}$ value is distinguished only 
for binary types when calculating $N_{\rm{fb}}/N_{\rm{b}}$ ratio in different $log~AL$ bin, 
e.g., $N_{\rm b}$ = 632 for detached binaries (listed in Table~3) as building the red 
dotted line in Figure.~\ref{fig3}(b). Such a ratio measurement ($N_{\rm{fb}}/N_{\rm{b}}$)
excludes the effect of difference of the original $N_{\rm b}$, therefore, it indicates more clearly 
the flare-binary frequency in different binary types. Two peaks show 
up again in Figure.~\ref{fig3}(b) at the same $log~AL$ positions as in Figure.~\ref{fig3}(a). 
Semi-detached binaries (the green dotted line) have much larger $N_{\rm{fb}}/N_{\rm{b}}$ ratio 
compared to the other types. }

The actual number of binary systems ($N_{\rm b}$) in our working sample and the
ratio of $N_{\rm{fb}}/N_{\rm b}$ in different morphologies are listed in Table~3, which shows
that the detached and semi-detached binaries occupy 78\% of the entire working 
sample (column~2), 86\% of the flare binaries are from these two types (column~3), 
and the $N_{\rm{fb}}/N_{\rm b}$ ratios in these two types are 10-20\% higher 
than those in over-contact and ellipsoidal binaries (column~4). 
Detailed discussion is presented in the following content.

\subsection{KIC 5952403 and KIC 11347875}
As all the flare candidates are finally confirmed by eye, the process provides us a unique 
opportunity to confront `peculiar' flare binaries.
We pay special attention on two of them, i.e., KIC~5952403 and KIC~11347875, to figure out
whether the flaring properties are related with the special features of the host systems. 

\begin{figure}[t]
\includegraphics{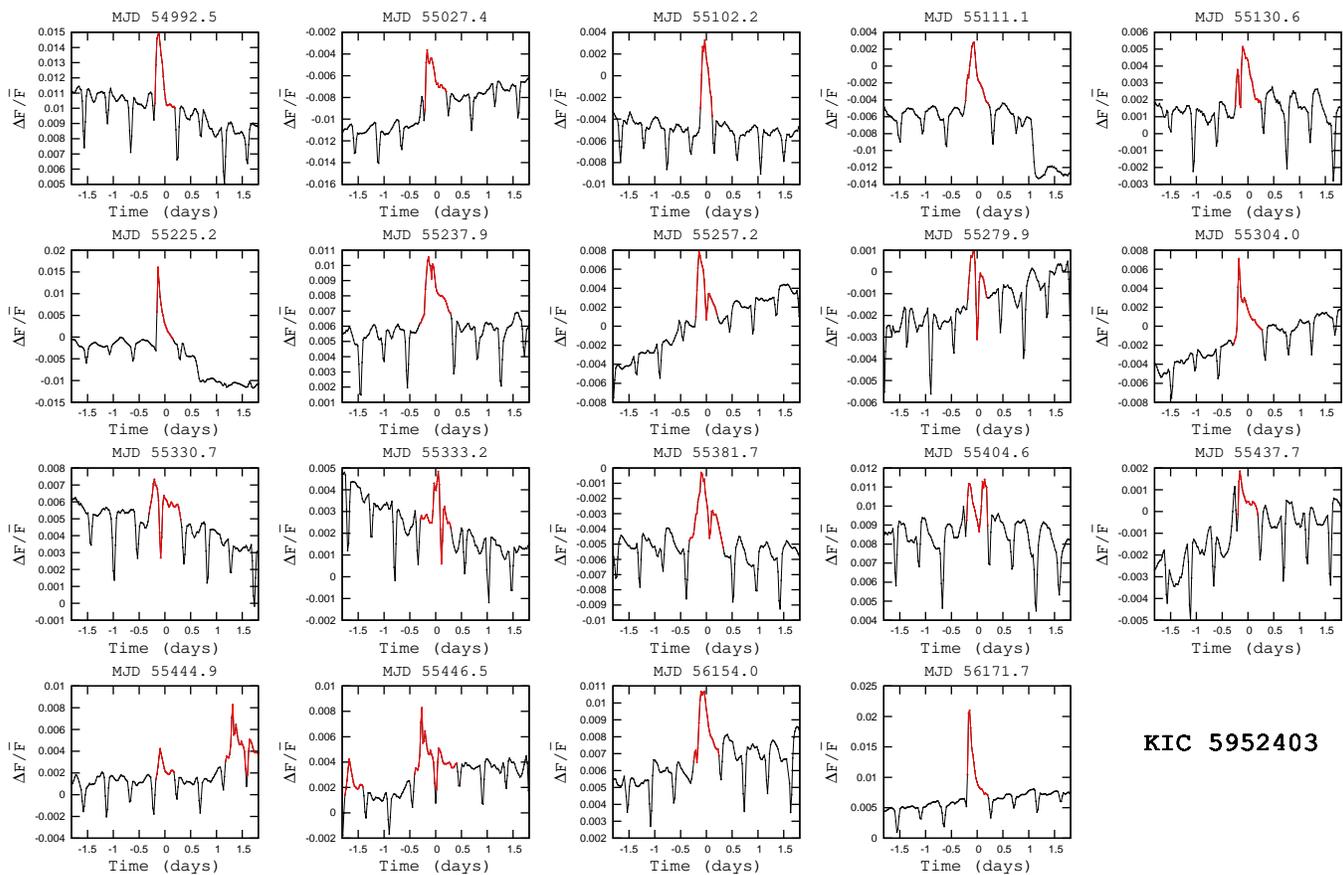}
\caption{\label{fig4} Light curves of 19 flare events detected on KIC 5952403. 
Flares are highlighted in red lines. The x-axis is the time (days) from the flare. 
The y-axis is the relative flux ($\Delta F/\bar F$) of the light curve.
\textbf{MJD is given on the top of each panel.} }
\end{figure}

KIC~5952403 (HD~181068) is a hierarchical triple system composed of an outer 
primary G-type red giant (we call it star `A') 
and an inner close binary with two late type red dwarfs 
(we call them star `B' and star `C', respectively). 
Fundamental parameters of the three components are listed in Tables~3 and 4 in \citet{Borkovits2013}.
We identify 19 flare events on the system from LC data. 
Light curves of the 19 flares are presented in Figure.~\ref{fig4}, in which the x- and y-axis 
\textbf{give the time (days) from the flare} and the relative flux ($\Delta F/\bar F$) 
of the light curve, respectively. All the flares are highlighted in red color in the figure. 

\begin{figure}[t]
\includegraphics{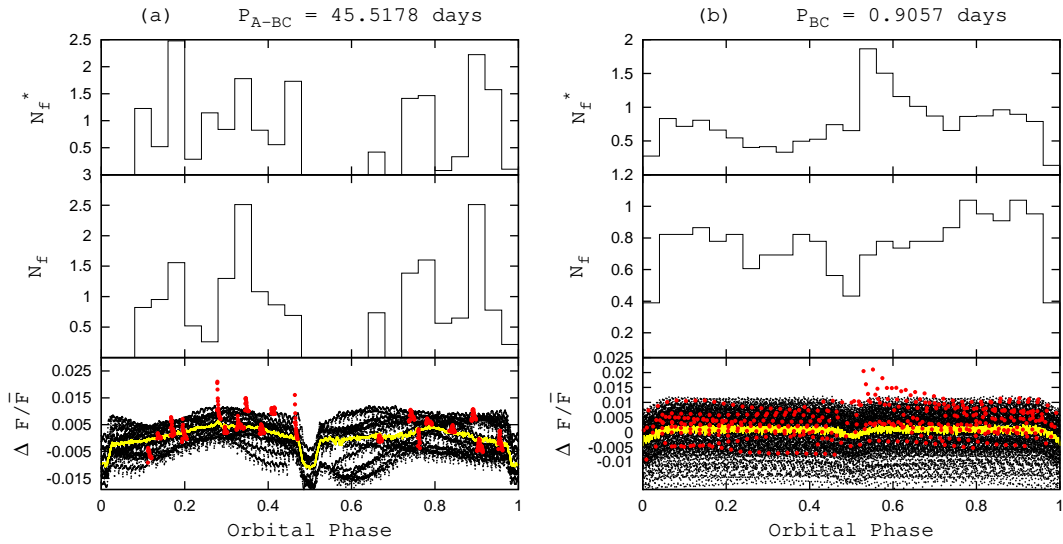}
\caption{\label{fig5} Flare number distribution of KIC~5952403 when the light curves are folded with 
$\bm{P_{\rm{A-BC}}}$ = 45.5178 days (panel a) and with $\bm{P_{\rm{BC}}}$ = 0.9057 days (panel b), respectively. 
Bottom panels: the black dots and the yellow lines present the orbital phase folded light curves of the 
primary and the robust mean values of the light curves, respectively; the rod 
dots are the observed flares. Top panels: amplitude-weighted flare number distributions ($N_{\rm f}^\star$)
along the orbital phase.
Middle panels: normal flare number distributions ($N_{\rm f}$) along the orbital phase.}
\end{figure}

We fold the LC light curve of KIC~5952403 into the triple-system orbital period 
$\bf{P_{\rm{\bf{A-BC}}}}$ = 45.5178~days (Figure.~\ref{fig5}(a)) and the inner dwarf pair orbital period 
$\bf{P_{\rm{\bf{BC}}}}$ = 0.9057~days (Figure.~\ref{fig5}(b)), respectively. $\bf{P_{\rm{\bf{A-BC}}}}$ and $\bf{P_{\rm{\bf{BC}}}}$ values 
are from the KEB catalog.
In Figure.~\ref{fig5}, the flare-amplitude weighted flare number distributions ($N_{\rm f}^\star$
\footnote{\textbf{(a)$\sum_{i=1}^{bin}N_{\rm f_i}=N_{\rm f}$; 
(b)Const $\times \sum_{i=1}^{bin} (N_{\rm {f_i}} \times Amp_{\rm i})=N_{\rm f}$; 
(c)$Amp_i=(\Delta F/\bar F)_{flare_{\rm max}}$, where $flare_{\rm max}$ 
is the highest-energy flare in the $ith$ phase bin;
(d)$N_{\rm f_i}^\star = N_{\rm f_i} \times Amp_{\rm i} / Const$.}}) are presented in the top panels, 
and the normal flare number distributions ($N_{\rm f}$) are given in the middle panels. 
The black dots and yellow lines in the bottom panels represent the orbital-phase folded light 
curves and the robust mean values of the light curves, respectively.
\textbf{$N_{\rm f}^\star$ is weighted by the relative amplitude ($\Delta F/\bar F$) of the highest-energy 
flare in each phase bin. The detailed description is given in mathematic functions in the footnote.}

\citet{Derekas2011} found that almost all flares on the system appear right after 
the shallower minimum of the inner dwarf pair (as shown in Figure.~\ref{fig4}),
and there is a clear $N_{\rm f}^\star$ enhancement at phase 0.55 in Figure.~\ref{fig5}(b), 
both features suggest that flaring activity seems to be associated with the close pair
\citep{Derekas2011}. 
However, most of the flares last around half day from Figure.~\ref{fig4} display, 
and some of them even last longer than 1 day. 
For a triple system, if the duration of a flare is longer than the orbital period of the 
inner pair, the flare is more likely associated with star~`A' rather than the inner pair,
because otherwise, the flare may not stay visible in a full orbital period without eclipse,
especially for an edge-on system. 

One flare is actually observed on KIC~5952403 
during the secondary eclipse from the {\it Kepler} SC data  \citep{Czesla2014}. 
It does not appear at phase 0.5 in our Figure.~\ref{fig5}(a) as we use LC data in the work. 
The flare occurs when the dwarf pair is hidden behind the giant, 
it is thus definitely ascribed to the red giant. 
Totally 7 flares are reported by \citet{Czesla2014} in the $\it{Kepler}$ SC light curve of 
KIC~5952403.
The authors argued that all these flares should be originated on the red giant, 
because the peak flare luminosity they calculated is comparable to or even higher than 
the total luminosity of the dwarf pair, or,
it would require an unreasonably large region on the dwarfs to reproduce such flare peak. 
 
Following the same discussion as in section~5.4.1 in \citet{Czesla2014}, 
Table~4 gives properties of the 19 flares identified in the $\it{Kepler}$ LC data in this work. 
According to the column~2 information in Table~4, the peak luminosities of the 19 flares 
range between 0.22\% and 1.72\% of the giant's luminosity 
($L_{\rm{bol\_giant}}\sim$92.8$\pm$7.6$L_{\sun}$, Table~1 in  \citet{Czesla2014}), 
which refers to 0.2 to 1.6 $L_{\rm{\sun}}$.  
Assuming that the flaring material seen by $\it{Kepler}$ has a temperature of $T_{\rm{eff,f}}$ = 10,000~K, 
and the dwarf stars have radii of $R_{\rm s}\approx$0.8~$R_{\sun}$ ( \citet{Czesla2014} and references therein), 
and using the formula of $\small{f=2\frac{L_{\rm{peak}}}{\sigma T_{\rm{eff,f}}^4 4\pi R_{\rm s}^2}}$, 
we estimate that between 6\% and 53\% of the visible hemisphere of either dwarf star has to be 
covered by flaring material in order to produce such flare peaks, which is too much larger 
than current observation evidence, e.g., an unusually intense flare observed on EV~Lac star 
requires $\approx$3\% hemisphere coverage \citep{Osten2010}. 
In contrast, it requires only 0.03-0.19\% coverage if the flares occur on the giant. 
It seems that
the observation from this work supports the conclusion in \citet{Czesla2014} that the giant 
is more likely the birthplace of all the flares. 

\begin{figure}[t]
\includegraphics{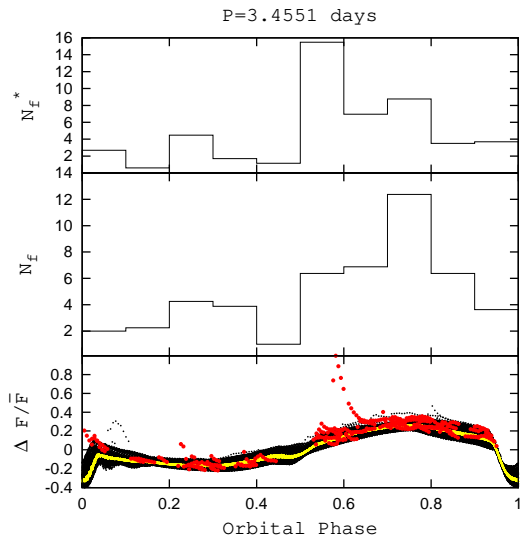}
\caption{\label{fig6} Flare number distributions and light curve of KIC~11347875.}
\end{figure}

KIC~11347875 is a binary system. Its orbital period is $P$ = 3.4551~days (from KEB catalog).
Both KIC parameters \textbf{($\bf{T_{\rm{eff}}}$ = 4656~K, $log~g$ = 2.581) and the study of \citet{Armstrong2014} indicate that the system contains two late type red giants}. 
\textbf{The $T_{\rm eff}$ and $log g$ parameters are extracted from the {\it Kepler} input 
catalog\footnote{http://archive.stsci.edu/kepler/stellar17/search.php}.}
Figure.~\ref{fig6} gives the light curve and the flare number distributions of KIC~11347875. 
There are 49 flares detected on the system. 
These flares (red dots in the bottom panel in Figure.~\ref{fig6}) are more frequently captured 
around phase 0.7-0.8. 
In the bottom panel in Figure.~\ref{fig6}, both observed light curve 
(black dots) and their mean values (yellow lines) are in asymmetric shape against 
phase 0.5, and the shape stays almost invariable during the entire observing time (over 1500 days). 
If the asymmetric shape and the flare generation are related to the active stellar spot region, 
then it implies that the movement of the group of spots is synchronized with the orbital motion, 
and the group of spots may cause the non-axisymmetry distribution of the magnetic fields in 
the system \citep{Berdyugina1998a}.

\subsection{Effect of Morphology}

\begin{figure}[t]
\includegraphics{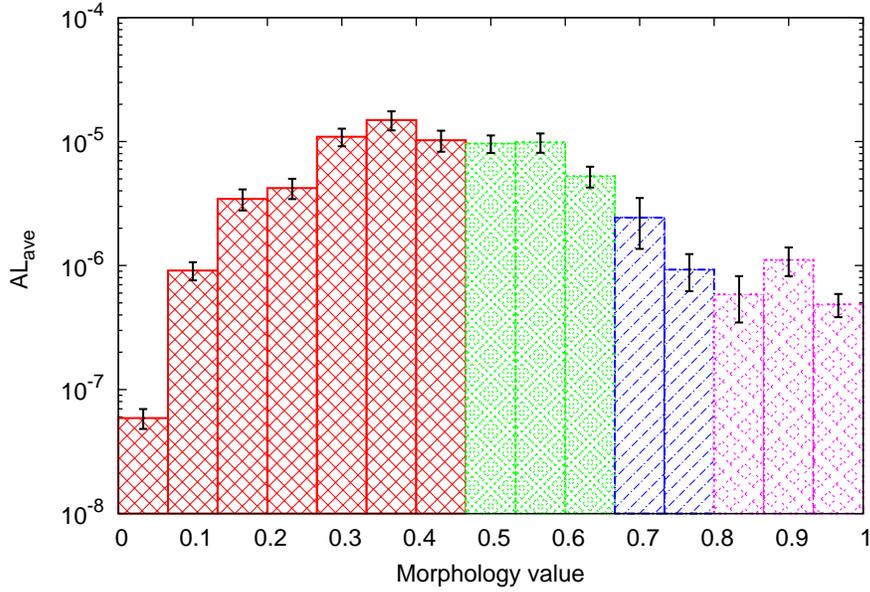}
\caption{\label{fig7} \textbf{Distribution of $AL_{\rm{ave}}$ along morphology values. 
The red, green, blue and magenta colors represent the distributions of detached, 
semi-detached, over-contact and ellipsoidal binaries, respectively.
The error bars represent 1$\sigma$ uncertainty in $AL$ calculations and the square root of event 
numbers in each morphology-value bin. }}
\end{figure}

Figure.~\ref{fig7} shows the $AL_{\rm{ave}}$\footnote{\textbf{$\bm{AL_{\rm{ave}}}=\sum AL/N_{\rm b}$, 
where $\sum AL$ sums up $AL$ of all the flares in a given parameter region, 
and $N_{b}$ is the number of binaries in the same region.}}
variation along morphology values for the total $\textbf{1049}$ binaries.
As presented in the figure, flare activities are more frequently generated on the 
detached ($Mor<$~0.5) and semi-detached binaries (0.5~$<Mor<$~0.7) compared to the contact systems.  
We also notice that the $AL_{\rm ave}$ variation is related to the interacting activities between 
two components in a binary, more specifically,
$AL_{\rm ave}$ increases monotonically with the increasing of morphology values for detached binaries, 
$AL_{\rm ave}$ decreases with the increasing of morphology values for semi-detached and over 
contact binaries, and $AL_{\rm ave}$ keeps roughly constant for ellipsoidal binaries. 
$Mor>$~0.5 means at least one of the components has filled its Roche lobe and mass transfer begins. 
The two components become closer as $Mor$ is larger. 
When $Mor>$~0.7, two stars in a binary are so close and start sharing a common envelope, 
and the $AL_{\rm ave}$ variation starts becoming invisible in this condition. 
The $AL_{\rm ave}$ of over contact binaries ($Mor>$~0.7) is depressed compared
to those fast rotating detached binaries ($Mor\sim$~0.5),
which implies that the magnetic activity in common enveloped binaries may be frustrated
by the development of the common envelope \citep{Qian2001,Qian2003}.
\textbf{Error bars in Figure.~\ref{fig7} represent 1$\sigma$ uncertainty in $AL$ 
measurements and the square root of $N_{\rm b}$ number in each morphology-value bin 
(the method described in Figure.~2 and 3 in \citet{Maehara2012}). 
\citet{Maehara2012} studied superflares on solar-type stars with $\it{Kepler}$ data, 
where they estimated that the uncertainties in luminosities and energies of flares are $\pm$60\%.
We use the $\pm$60\% uncertainties in our work when calculating $EW$ values in eq.~(2), 
based on mainly two reasons: (1) the duration time of most of the flares we detected is comparable 
to that of superflares (typically a few hours), and the comparability may extend 
to stellar energy; 
and (2) it is impossible to calculate the reliable luminosity (or energy) uncertainty for binary systems, 
because the reliability of all the essential parameters, such as effective temperature, surface gravity and radius, are low as they are estimated by using the mixed light of a binary system.}

\subsection{The Effect of Orbital Period and Rotation}

Orbital period ($\bm{P_{\rm{orb}}}$) is one of the decisive parameters to influence the evolution 
of binary systems. We know that shorter period usually refers to faster rotation due to 
tidal interaction \citep{Walter1981}, 
and thus implies the increased magnetic activity \citep{Noyes1984}. 
For instance, the RS~CVn variables on which flares are observed are short period detached 
binaries with strong magnetic activity \citep{Berdyugina2005}. 

Figure.~\ref{fig8}(a) is the $\bm{AL_{\rm{ave}}}$ variation along $\bm{P_{\rm{orb}}}$ which is extracted from the KEB catalog.
$\bm{AL_{\rm{ave}}}$ is increasing with shortening $\bm{P_{\rm{orb}}}$ up to a critical value at $\bm{P_{\rm{orb}}}\sim$3~days, 
thereafter, the $\bm{AL_{\rm{ave}}}$ strength decreases with the decreasing $\bm{P_{\rm{orb}}}$.
$\bm{AL_{\rm{ave}}}$ shows no correlation with $\bm{P_{\rm{orb}}}$ when $\bm{P_{\rm{orb}}}>$35~days. 
Figure.~\ref{fig8}(b) is the morphology distribution along the $\bm{P_{\rm{orb}}}$. 
All the contact systems ($Mor>$~0.7) shrink in a short orbital period 
range of $\bm{P_{\rm{orb}}}<$~10~days. 
Most of the binaries with $\bm{P_{\rm{orb}}}>$~35~days 
are quite detached systems shrinking in a narrow morphology range of $Mor<$~0.2.

The rotation rate of a star plays also an important role in flaring rate \citep{Nielsen2013}. 
For instance, \citet{Notsu2013} and \citet{Shibayama2013} find that the flare frequency 
is lower in slowly-rotating solar-type stars. For G, K and M stars, \citet{Candelaresi2014} 
find that flaring rate is increasing with rotation rate up to a critical value, and then, 
the flaring rate decreases linearly with increasing rotation rate. The same feature also fits 
for the binary systems. Figure.~\ref{fig9}(a) shows the $\bm{AL_{\rm{ave}}}$ variation along the rotation period ($\bm{P_{\rm{rot}}}$), 
where the $\bm{AL_{\rm{ave}}}$ increases with the increasing rotation rate up to $\bm{P_{\rm{rot}}}\sim$1.5~days, 
and then, the $\bm{AL_{\rm{ave}}}$ starts decreasing with the decreasing $\bm{P_{\rm{rot}}}$. 
$\bm{P_{\rm{rot}}}$ is only calculated for detached binaries in the work.
We follow the studies such as \citet{Walkowicz2011}, \citet{Maehara2012} and \citet{Nielsen2013}, 
assuming that the period of brightness modulation corresponds to the rotation period of 
a star, and then we search the out of eclipse parts of binary light curves using Lomb-Scargle
periodogram to measure the stellar rotation period of the primary.
\textbf{We compute a Lomb-Scargle periodogram for each target in each quarter, and then 
we identify the peak of maximum power and record its period.
Median value of the recorded periods of all the quarters is determined as the rotation period.} 
Figure.~\ref{fig9}(b) shows the correlation between $\bm{P_{\rm{orb}}}$ and $\bm{P_{\rm{rot}}}$ for detached systems.
The monotonically increasing correlation between two parameters stops at $\bm{P_{\rm{rot}}}\sim$32~days.

\subsection{Effect of Orbital Phase}
It has been suggested that flares on active close binaries could arise due to magnetic reconnection 
field lines connecting two components \citep{Simon1980, van1988, Gunn1997},
\textbf{for instance, tidal effect can play important role for the enhanced magnetic activities in 
RS~CVn stars \citep{Cuntz2000}.}
As the gravitational interaction and mass transfer between two stars may break the spherical 
symmetry of stars and influence the development of magnetic fields, 
the density of magnetic field lines around the binary could present distribution 
related with the orbital phase \citep{Holzwarth2003a,Holzwarth2003b}, and thus the flaring 
rate may be also sensitive to the orbital phase.
Geometrically, a binary system presents its largest cross-section on line of sight at orbital phase 
0.25 and 0.75 (primary eclipse at phase 0). If the magnetic activity is randomly distributed 
on the surface of both components, the $N_{\rm f}$ distribution should be modified simply due to  
the eclipse, i.e., the $N_{\rm f}$ minimum at phase 0 and 0.5, and the flat $N_{\rm f}$ distribution at other 
phases; and meanwhile, any `peculiar' $N_{\rm f}$ distribution would indicate the different magnetic 
features or different magnetic activity areas.
  
KIC~2309587 is a detached binary with $N_{\rm f}$=128 (Table~1), 
$T_1$=5166~K and $T_2$=5577~K \citep{Armstrong2014}. 
The value of \textbf{$\bf{log~g}$=4.326 (from revised KIC, \citet{Huber2014})} indicates both components are main-sequence stars. 
It is selected as an example to detect the correlation between flaring rate 
and orbital phase in the detached system. 
The amplitude-weighted flare number distribution ($N_{\rm f}^\star$; top panel), flare number distribution 
($N_{\rm f}$; middle panel) and orbital-phase folded light curves ($\Delta F/\bar F$; bottom panel) 
of the binary are presented in Figure.~\ref{fig10}(a), respectively. 
\textbf{Symmetric light curve in the bottom panel indicates the non-existent of large star-spots. }
Both $N_{\rm f}^\star$ and $N_{\rm f}$ drop at phase 0 and 0.5, and keep almost constant at other phases,
which indicates that flares are more likely originated on both components
in the system, and the flaring rate is irrelevant with the orbital phase.
 
Figure.~\ref{fig10}(b) is the same as Figure.~\ref{fig10}(a), but for KIC~11457191, 
which is a semi-detached binary with 210 flares detected (Table~1). 
\textbf{Revised KEB catalog shows the system has $\bm{T_{\rm{eff}}}$=5858~K, 
$\bf{log g}$=4.486 and orbital period of $\bm{P_{\rm{orb}}}$=2.298~days.}
The flare-number-distribution variation of the binary is still dominated by 
the eclipse between two components. As mass transfer begins in the system, and $N_{\rm f}$ 
keeps roughly flat at phase $>$ 0.5, flaring should be more active on the obtaining-mass 
secondary, but still, no sign of the correlation between magnetic activity and 
orbital phase is presented in Figure.~\ref{fig10}(b).

We use \textbf{20} non-eclipse binaries with $Mor>$~0.7 to examine the correlation. 
Totally \textbf{203} flares are distributed along the orbital phase in Figure.~\ref{fig11}(a) and (b). 
Figure.~\ref{fig11}(c) gives the orbital-period folded light curve of KIC~11153627 as an example 
to illustrate the typical baseline-flux variation of an ellipsoidal system. 
In both Figure.~\ref{fig11}(a) and (b), the significant enhancement of flare numbers amazingly 
shows up at around phase 0.25 and 0.75, which is consistent with positions of the largest 
amplitudes in Figure.~\ref{fig11}(c).  
We know that even for the non-eclipse binaries, the cross-section on line of sight still varies 
with orbital phase due to tidal distortion, so does the surface luminosity, and therefore, 
the maximum light-curve amplitude of each binary can be viewed as the upper limit of 
the cross-section discrepancy, which is approximately 12\% for KIC~11153627 
(as shown in Figure.~\ref{fig11}(c)), in other words, 
the changes of cross section should be no more than 12 percent for the binary.

The averaged cross-section variation for the \textbf{20} non-eclipse samples is 
actually 2.9$\pm$2.6\%. 2.6\% is the 1-$\sigma$ deviation, and
thus, the 12 percent can be approximately considered as the upper limit 
of the cross-section modification. 
\textbf{In order to investigate how essential the cross-section is on the flare detection, we assume
the minimum cross section of a binary is 1 at phase 0 and 1, and the maximum cross section is 
1.12 at phase 0.25 and 0.75 as the maximum cross-section variation is 12\%, 
and then we calculate the flare numbers again by dividing the numbers with the corresponding 
cross-section area. The red dashed line and the black solid line in Figure.~11 present 
flare number distributions with and without cross-section correction, respectively, and both 
lines in $N_{\rm f}^\star$ and $N_{\rm f}$ distributions in Figure.~11 show clear enhancement 
at phase 0.25 and 0.75. Such variation can be larger than what is expected from the 
difference of cross section between phase 0.25(0.75) and phase 0(1).}

\section{Discussion and Conclusion}
In this paper, we present the results of flare-binary identification from $\it{Kepler}$ LC data.
We identify \textbf{234} flare binaries out of the \textbf{1049} binary samples in the KEB catalog.
We use 4-terms Fourier series to capture flare features and finally \textbf{6818} flares are 
detected on the \textbf{234} flare binaries. 

We pay special attention on a triple system KIC~5952403, also known as HD~181068.
Previous studies argue that flares on the system should be generated simply  
on the inner pair \citep{Derekas2011} or simply on the outer giant \citep{Czesla2014}.
By estimating the necessary hemisphere coverage according to the observed flare peak luminosity, 
we confirm that the 19 flares detected in this work from the $\it{Keple}$ LC light curve 
should be originated on the giant star. 

We compare the flare-binary fraction in binaries of different morphology types. 
The result shows that the fractions in over-contact and ellipsoidal binaries are 
approximately 10-20\% lower than those in detached and semi-detached systems. 
Contact binaries are usually more luminous compared to detached ones, and high luminosity 
naturally decreases the visibility of flare events, which the 10-20\% difference should be 
mainly attributed to. 
In addition, the existence of common envelops in contact systems may depress 
the activity of magnetic field and thus influence the flare generation.

\begin{figure}[t]
\includegraphics{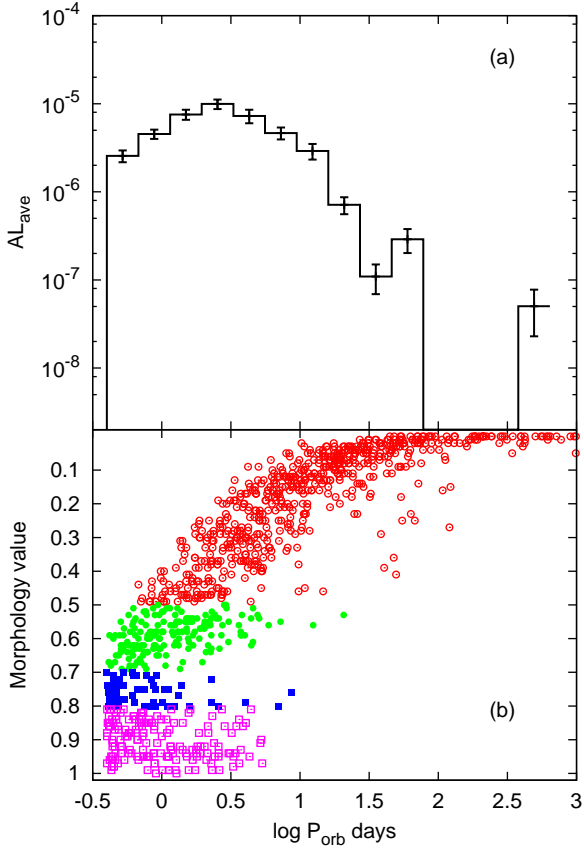}
\caption{\label{fig8} Panel~(a): $\bm{AL_{\rm{ave}}}$ variation along orbital period.
\textbf{Panel~(b): Correlation between morphology value and orbital period. 
The red open circles, green solid circles, blue solid squares and magenta open squares 
represent detached, semi-detached, over-contact and ellipsoidal binaries, respectively.
The error bars represent 1$\sigma$ uncertainty in $AL$ calculations and the square 
root of event numbers in each orbital-period bin.}}
\end{figure}

\begin{figure}[t]
\includegraphics{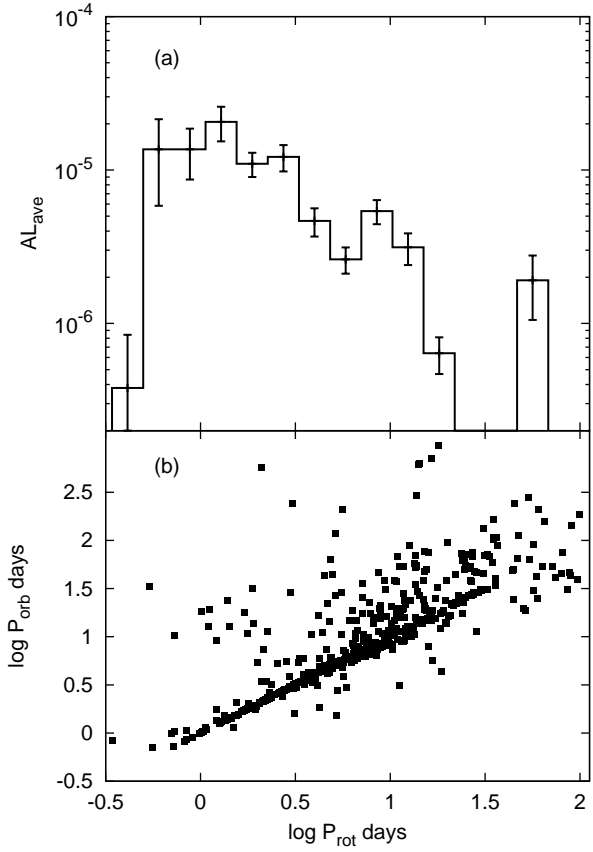}
\caption{\label{fig9} Panel~(a): $\bm{AL_{\rm{ave}}}$ variation along rotation period.
Panel~(b): Correlation between orbital period ($\bm{P_{\rm{orb}}}$) and rotation period 
($\bm{P_{\rm{rot}}}$), and the correlation is discussed for only detached binaries.
\textbf{The error bars represent 1$\sigma$ uncertainty in $AL$ 
calculations and the square root of event numbers in each rotation-period bin. }}
\end{figure}

\begin{figure}[t]
\includegraphics{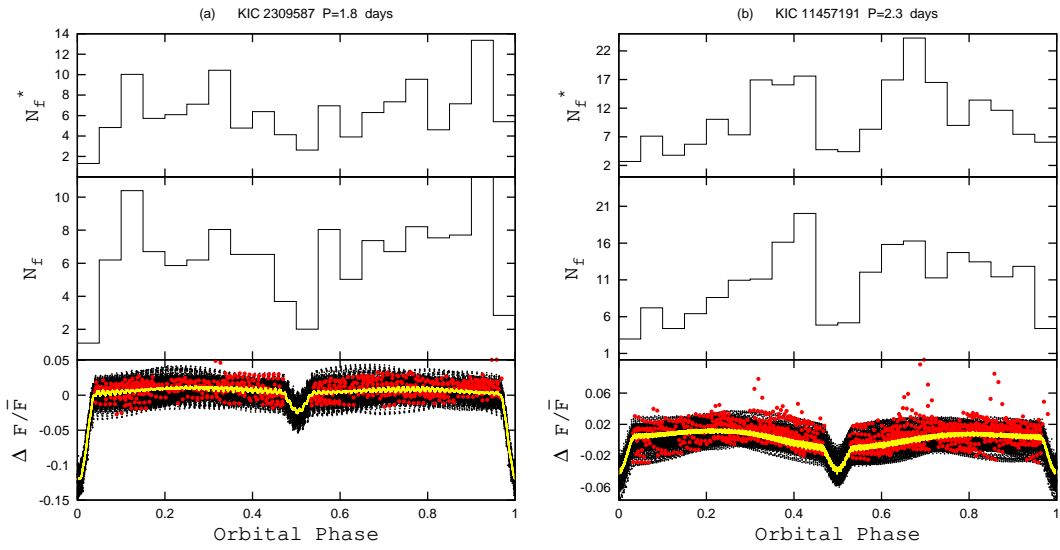}
\caption{\label{fig10} Flare number distributions and light curves of KIC~2309587 (panel a) 
and KIC~11457191 (panel b), respectively.}
\end{figure}

We identify a parameter named the flare `activity level' ($AL$, eq.(3)) to estimate the ratio of 
flare energy over the baseline energy in a binary. We discuss the $\bm{AL_{\rm{ave}}}$ variations 
along the $\bm{P_{\rm{orb}}}$ and $\bm{P_{\rm{rot}}}$ (calculated for only detached binaries) parameters. 
We report that $\bm{AL_{\rm{ave}}}$ increases with decreasing $\bm{P_{\rm{orb}}}$ and/or $\bm{P_{\rm{rot}}}$ up 
to the critical values at $\bm{P_{\rm{orb}}}\sim$3~days or $\bm{P_{\rm{rot}}}\sim$1.5~days, 
thereafter, the $\bm{AL_{\rm{ave}}}$ starts decreasing no matter how fast the stars rotate. 
When considering the $\bm{AL_{\rm{ave}}}$ variation in different morphology types, we find that
the variation is related to the interacting activities between two stars in a binary. 
As presented in Figure.~\ref{fig8}, when $Mor>$~0.5, meaning at least one of the components 
has filled its Roche lobe and mass transfer begins, $\bm{AL_{\rm{ave}}}$ stars decreasing with 
the increasing $Mor$; when $Mor>$~0.7, meaning the two stars in a binary are so close and 
start sharing a common envelope, $\bm{AL_{\rm{ave}}}$ variation stars becoming invisible in this condition. 

\begin{figure}[t]
\includegraphics{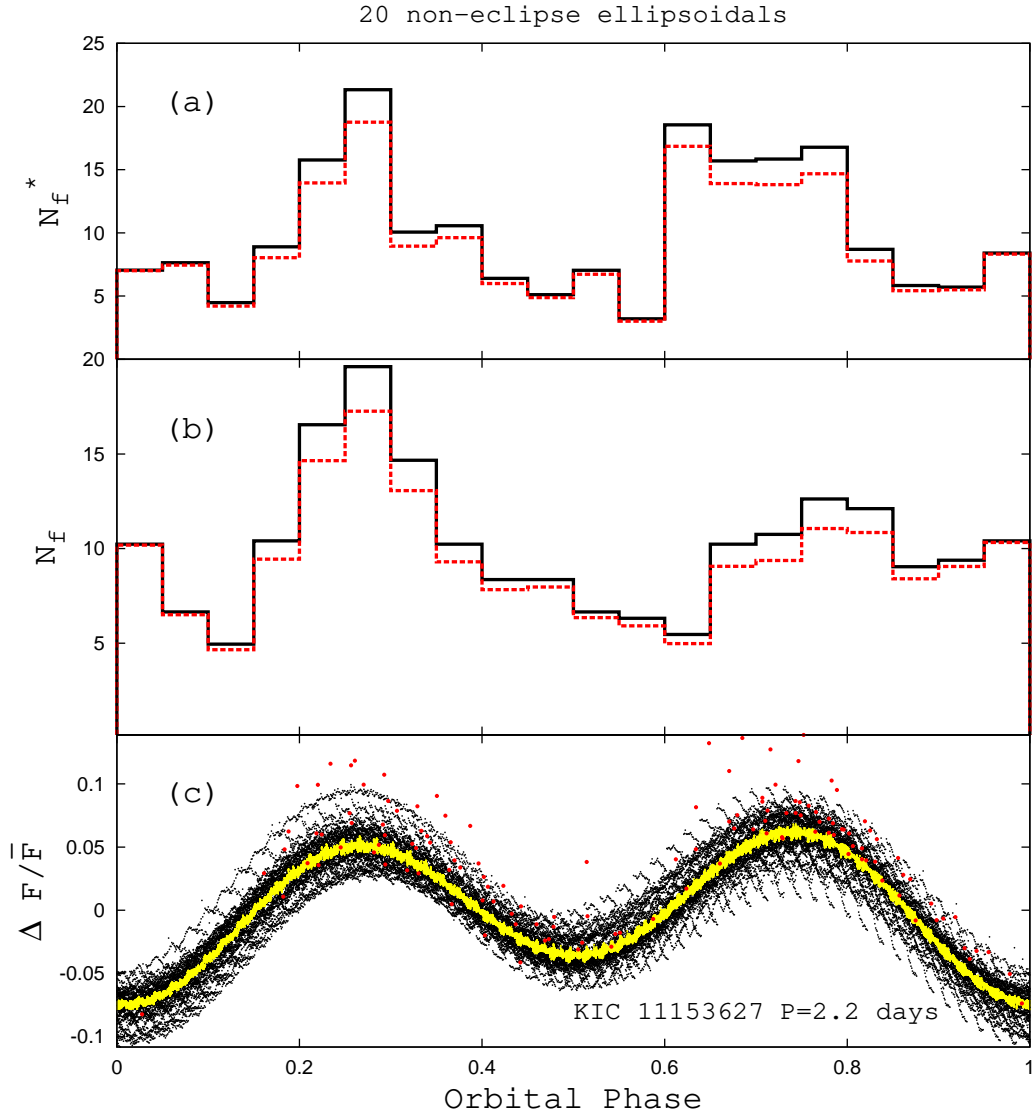}
\caption{\label{fig11} \textbf {Flare number distributions of \textbf{20} non-eclipse ellipsoidal binaries. 
Panel~(a): amplitude weighted flare number distribution ($\bm{N_{\rm f}^\star}$) along the orbital phase.
Panel~(b): normal flare number distribution ($\bm{N_{\rm f}}$) along the orbital phase.
Panel~(c): orbital-period folded light curve of KIC~11153627. 
In both Panel~(a) and (b), the red dashed lines represent the distributions after the correction of
 cross-section on line of sight, and the black solid lines are the corresponding distributions without 
 cross-section correction.}}
\end{figure}

We examine the flaring rate as a function of orbital phase in 2 eclipsing binaries 
on which a large number of flares are detected.
It appears that there is no correlation between flaring rate and orbital phase in these 2 binaries. 
In contrast, when we examine the function with \textbf{203} flares on \textbf{20} non-eclipse ellipsoidal binaries, 
bimodal distributions of both $N_{\rm f}^\star$ and $N_{\rm f}$ show up at orbital phase 0.25 and 0.75.
We argue that the maximum 12 percent cross-section variation can not afford the observed 
flare-number difference between phase 0.25(0.75) and 0(0.5), as presented in Figure.~\ref{fig11}(a) and (b). 
We conclude that the bimodal distribution might be still related with the slight orbital 
inclination and/or gravitational distortion, and meanwhile,
\textbf{such variation can be larger than what is expected from the 
difference of cross section between phase 0.25(0.75) and
phase 0(1).}

\acknowledgements 
QG is grateful to Yu Bai and Song Wang for useful suggestions.
YX thanks the 973 Program 2014CB845700. JFL acknowledges the 973 Program 2014CB845705.
SG is supported by the Fundamental Research Funds for the 
Central Universities.
The paper includes data collected by the {\it Kepler} mission.
Funding for the {\it Kepler} mission is provided by the NASA Science Mission Directorate.
All of the data presented in this paper were obtained from the 
Mikulsk Archieve for Space Telescope (MAST). STScI is operated by the Association 
of Universities for Research in Astronomy, Inc., under NASA contract NAS5-26555.
Support for MAST for non-HST data is provided by the NASA Office of Space 
Science via grant NNX09AF08G and by other grants and contracts.

\bibliography{ref}

\clearpage
\oddsidemargin=-2cm
\topmargin = -2cm

\begin{table}

\centering
\caption{\label{table1} Fundamental parameters of 234 flare binaries. 
The KIC number, effective temperature ($\bm{T_{\rm{eff}}}$), surface gravity ($log~g$), orbital period ($\bm{P_{\rm{orb}}}$)
and morphology value ($Mor$) are extracted from the KEB cataloge. 
The light curve amplitude ($Amp$), flare activity level ($AL$),
flare frequency ($f$(flares/day)), total number of flares ($N_{\rm f}$) and the peak relative flux of flares 
($log (\Delta F_{\rm{max}}/\bar F)$) are calculated in the work. 
$\Delta F_{\rm{max}}$ is calculated by substituting $F$ of the highest energy flare into eq.~(1).
\textbf{$DR_{\rm notes}$ is the LC-PDC data release version we adopt for each flare binary.}
The complete table is available electronically.}

\begin{tabular}{l l l l l l l l l c l}
\noalign{\smallskip}
\hline
KIC        &  $T_{\rm{eff}}$[K] & $log~g$ & $P_{\rm{orb}}$[day] & $Mor$ & $Amp$  & $AL$          & $f$[1/day]      & $N_f$   & $log (\Delta F_{\rm{max}}/\bar F)$ & $DR_{\rm notes}$\\ 
\hline 
002305372  &  5664  &  3.974  & 1.405   &  0.58  &  0.066  &  1.232e-05  & 0.015  & 20  & -1.043  & 24 \\  
002447893  &  5059  &  4.447  & 0.662   &  0.58  &  0.043  &  1.415e-05  & 0.016  & 21  & -1.296  & 24 \\ 
002556127  &  5920  &  4.629  & 0.419   &  0.59  &  0.003  &  5.365e-08  & 0.001  & 1  & -2.583  & 24  \\ 
002557430  &  6248  &  4.103  & 1.298   &  0.51  &  0.010  &  6.314e-06  & 0.060  & 81  & -2.040 & 24 \\ 
002569494  &  5114  &  4.471  & 1.523   &  0.58  &  0.186  &  8.435e-05  & 0.058  & 25  & -1.264  & 24  \\ 
002577756  &  5630  &  4.953  & 0.870   &  0.63  &  0.094  &  2.946e-06  & 0.007  & 10  & -1.775 & 24 \\ 
002835289  &  6228  &  4.265  & 0.858   &  0.92  &  0.023  &  3.431e-06  & 0.011  & 15  & -1.584 & 24\\ 
003114667  &  ---  &  ---  & 0.889   &  0.52  &  0.041  &  6.585e-05  & 0.024  & 15  & -0.873   & 24 \\ 
003218683  &  5119  &  4.659  & 0.772   &  0.69  &  0.173  &  2.602e-05  & 0.007  & 10  & -0.856 & 24\\ 
003338660  &  5722  &  4.276  & 1.873   &  0.60  &  0.077  &  1.904e-05  & 0.018  & 14  & -1.031 & 24 \\ 
\hline
\end{tabular}
\end{table}

\begin{table}

\centering
\caption{\label{table2} Parameters of the \textbf{6818} flares.}

\begin{tabular}{l l l l l}
\noalign{\smallskip}
\hline
KIC  &  $\bm{t_{\rm max}}$[day] & $EW$[day]  &  $\Delta F/\bar F$ &$\Delta t$[hr] \\
\hline
002305372 &5143.30545 &0.000140 &0.003621 & 1.471141   \\
002305372 &5159.56975 &0.000277 &0.008178 &1.961526   \\
002305372 &5166.43511 &0.000086 &0.001994 &1.471150\\
002305372 &5297.29229 &0.000299 &0.005592 &1.961720\\
002305372 &5422.18660 &0.001007 &0.012124 &4.904034\\
002305372 &5446.23655 &0.001099 &0.043413 &1.471182\\
002305372 &5458.53721 &0.000930 &0.011786 &4.413509\\
002305372 &5511.41691 &0.001581 &0.019670 &4.413426\\
\hline

\end{tabular}
\tablecomments{The columns are: (1) the KIC number, (2) time of flare maximum ($t_{\rm{max}}$)
relative to BJD~2450000, (3)~$EW$ (days), (4) relative flux amplitude ($\Delta F/\bar F$),
(5) flare duration ($\Delta t$ (hours)). 
The complete table and the light curves are available electronically.}
\end{table}

\begin{table}

\centering
\caption{\label{table3} Percentage of flare binaries in different 
morphology types.}

\begin{tabular}{l c c c}
\noalign{\smallskip}
\hline
Morphology &  $N_{b}$ & $N_{fb}$ & $N_{fb}/N_b$(\%) \\
\hline
Detached & 632 & 140  & 22.15\\
Semi-detached & 188 & 62 & 32.98\\
Over-contact & 59 & 7 & 11.86 \\
Ellipsoidal & 170 & 25 & 14.71 \\
\hline
All & 1049 & 234$^\star$  &22.31 \\
\hline

\end{tabular}
\tablecomments{The columns are: (1) morphology type of binary system, 
(2) number of binaries, (3) number of flare binaries, 
(4) the ratio of flare binaries in the same morphology. }
\tablecomments{$\star$ We identify totally \textbf{234} flare binaries from
the \textbf{1049} binary samples. The two peculiar systems (KIC~5952403 and KIC~11347875)
discussed in Section~3.1 are excluded from Figure~3 and Table~3 to avoid 
the infusion of atypical flare features.}
\end{table}

\begin{table}
\centering
\caption{\label{table4} Properties of the 19 flares in the $\it{Kepler}$ 
long-cadence light curve of KIC~5952403.}

\begin{tabular}{c l l c c}
\noalign{\smallskip}
\hline
\hline
$T_{\rm{peak}}^a$ & $f_{\rm{peak}}^b$ & $t_{\rm{rise}}^c$ & $t_{\rm{decay}}^c$ & $E_{\rm{flare}}^d$ \\
(d) & (\%) & (min) & (min)  & (erg) \\
\hline
159.916    &    0.47  &  59   &   221  &   2e+38 \\
194.736    &    0.60  &  29   &   265  &   3e+38 \\
269.685    &    0.81  &  59   &   103  &   3e+38 \\
278.553    &    0.81  &  118  &   206  &   4e+38 \\
297.984    &    0.38  &  103  &   250  &   2e+38 \\
392.610    &    1.72  &  15   &   206  &   4e+38 \\
405.217    &    0.45  &  118  &   309  &   3e+38 \\
424.589    &    0.80  &  44   &   250  &   4e+38 \\
447.312    &    0.25  &  88   &   177  &   1e+38 \\
471.322    &    0.87  &  74   &   324  &   3e+38 \\
497.969    &    0.22  &  74   &   368  &   2e+38 \\
500.748    &    0.26  &  250  &   177  &   1e+38 \\
549.095    &    0.50  &  162  &   309  &   3e+38 \\
572.267    &    0.26  &  250  &   44   &   1e+38 \\
605.022    &    0.26  &  29   &   250  &   1e+38 \\
612.337    &    0.27  &  59   &   191  &   8e+37 \\
613.747    &    0.57  &  103  &   485  &   3e+38 \\
1321.424   &    0.42  &  103  &   250  &   2e+38 \\
1339.058   &    1.46  &  29   &   250  &   4e+38 \\
\hline
\end{tabular}
\tablecomments{$^{(a)}$ Time of flare peak in BJD-2454833. 
$^{(b)}$ Flare maximum in fraction of stellar flux. 
$^{(c)}$ Duration of rise (between beginning and maximum) and 
decay (between maximum and end) phase of the flare. 
$^{(d)}$ Energy of the flare in white light ($\it{Kepler}$ bandpass).}
\end{table}

\clearpage

\end{document}